\documentclass[letterpaper]{article} 
\usepackage{aaai24}  
\usepackage{times}  
\usepackage{helvet}  
\usepackage{courier}  
\usepackage[hyphens]{url}  
\usepackage{graphicx} 
\usepackage{natbib}  
\usepackage{caption} 
\usepackage{algorithm}
\usepackage{algorithmic}
\usepackage{amsmath,amsfonts}
\usepackage{tabularx}
\usepackage{tabulary}
\usepackage{multirow}
\usepackage{makecell}
\usepackage{algorithm,algorithmic,paralist}
\usepackage[table, dvipsnames]{xcolor}

\usepackage{array,booktabs}
\usepackage{newfloat}
\usepackage{listings}
\DeclareCaptionStyle{ruled}{labelfont=normalfont,labelsep=colon,strut=off} 
\floatstyle{ruled}
\newfloat{listing}{tb}{lst}{}
\floatname{listing}{Listing}
\title{MTVHunter: Smart Contracts Vulnerability Detection Based on Multi-Teacher Knowledge Translation}
\author {
Guokai Sun\textsuperscript{\rm1}\equalcontrib, 
Yuan Zhuang\textsuperscript{\rm1}\equalcontrib\thanks{Corresponding author}, 
Shuo Zhang\textsuperscript{\rm1}, 
Xiaoyu Feng\textsuperscript{\rm1}, 
Zhenguang Liu\textsuperscript{\rm2,\rm3},
Liguo Zhang\textsuperscript{\rm1}
}
\affiliations {
    \textsuperscript{\rm 1} College of Computer Science and Technology, Harbin Engineering University, Heilongjiang, China\\
    \textsuperscript{\rm 2}The State Key Laboratory of Blockchain and Data Security, Zhejiang University, Hangzhou, China\\
    \textsuperscript{\rm 3}Hangzhou High-Tech Zone (Binjiang) Institute of Blockchain and Data Security, Hangzhou, China
}

\usepackage{bibentry}

\begin{document}

\maketitle

\begin{abstract}
Smart contracts, closely intertwined with cryptocurrency transactions, have sparked widespread concerns about considerable financial losses of security issues. To counteract this, a variety of tools have been developed to identify vulnerability in smart contract. However, they fail to overcome two challenges at the same time when faced with smart contract bytecode: (i) strong interference caused by enormous non-relevant instructions; (ii) missing semantics of bytecode due to incomplete data and control flow dependencies. In this paper, we propose a multi-teacher based bytecode vulnerability detection method, namely \textbf{M}ulti-\textbf{T}eacher \textbf{V}ulnerability \textbf{Hunter} (\textbf{MTVHunter}), which delivers effective denoising and missing semantic to bytecode under multi-teacher guidance. Specifically, we first propose an instruction denoising teacher to eliminate noise interference by abstract vulnerability pattern and further reflect in contract embeddings. 
Secondly, we design a novel semantic complementary teacher with neuron distillation, which effectively extracts necessary semantic from source code to replenish the bytecode.  
Particularly, the proposed neuron distillation accelerate this semantic filling by turning the knowledge transition into a regression task. 
We conduct experiments on 229,178 real-world smart contracts that concerns four types of common vulnerabilities. Extensive experiments show MTVHunter achieves significantly performance gains over state-of-the-art approaches. 

\end{abstract}
\section{Introduction}

In recent years, blockchain technology has gained widespread attention across various sectors, serving as a decentralized and transparent ledger system \cite{z1}. 
Smart contracts are Turing-complete programs running on top of the blockchain, such as Ethereum, controlling a massive amount of digital assets through arbitrary rules \cite{a8}.
However, the publicly available character of smart contracts poses serious security risks, allowing malicious attackers to exploit vulnerabilities for substantial gains, a typical example is DAO, which caused a loss of \$60 million ETH by Reentrancy vulnerability.
In addition, over \$150 million worth of ETH was frozen due to the Delegatecall vulnerability in 2017.
According to existing statistics \footnote{https://learnblockchain.cn/article/8644}, the cumulative amount of losses caused by smart contracts exceeded \$1.492 billion in the first half of 2024, which has an increase of 116.23\% compared to the same period in 2023. 

Existing tools of smart contract tend to consider source code to judge security issues due to its semantic richness and clarity. Unfortunately, smart contracts are frequently not open-sourced and only bytecode is accessible. 
Therefore, it is urgent to advance vulnerability detection on bytecode. Current approaches on bytecode are either based on fixed and labor-consuming expert rules \cite{a2} or employ label-target neural networks \cite{a7}. We scrutinized the released implementation of existing methods, and empirically observed that they suffer from two key problems: \textit{first}, they assign weights equally to all instructions rather than distinguishing critical instructions related to the specific vulnerability, smoothing their features in the detection process. 
\textit{Second}, a substantial obstacle in successfully detecting vulnerability on bytecode is the difficulty of effectively and completely exploit  control-flow and data-flow dependencies, as  the bytecode is inherently missing a portion of semantics.

These problems motivate us to come up with the following three key designs: 
1) We design a multi-teacher smart contract vulnerability detection framework that contains two teachers and one student, enabling a divide-and-conquer approach to these challenging problems.
2) To overcome the interference of vulnerability-irrelevant instructions (i.e., noise), we design an Instruction Denoising Teacher (IDT) with fine-defined abstract vulnerability patterns, which distinguish critical instructions and grant more weights in contract embeddings.
3) On the other hand, we propose a Semantic Complementary Teacher (SCT), which effectively extracts rich semantic knowledge from source code, and then maps to missing semantic features on bytecode via a simple but effective method, namely neuronal distillation, turning the knowledge translation into a neuronal regression task.
To validate and evaluate the proposed method, extensive experiments have been conducted on a dataset of 229,178 real-world smart contracts, and results unequivocally demonstrate that MTVHunter significantly outperforms the state-of-the-art approaches in smart contract vulnerability detection. Notably, we harness existing prevalent distillation methods on smart contracts vulnerability detection and compare their performance for the first time. The codes are available at https://github.com/KD-SCVD/MTVHunter.

To summarize, we make the following \textbf{key contributions:}
\begin{itemize}
	\item We propose a novel framework to detect vulnerability on bytecode, which is a multi-teacher network integrating instruction denoising and semantic complementary to a student network, respectively.
	\item We design abstract vulnerability patterns in IDT to eliminate the interference of none-relevant instructions and reflect in contract embeddings.
	\item We present a novel neuron distillation in SCT, effectively and completely restore the missing semantic from source code to bytecode.
	\item We validate and analyze MTVHunter by extensive comparisons on 229,178 real-world smart contracts and achieve state-of-the-art performance of  vulnerability  detection on bytecode.
	
\end{itemize}
\begin{figure*}[htbp]
	\centering
	\label{f1}
	
	\includegraphics[width=\linewidth]{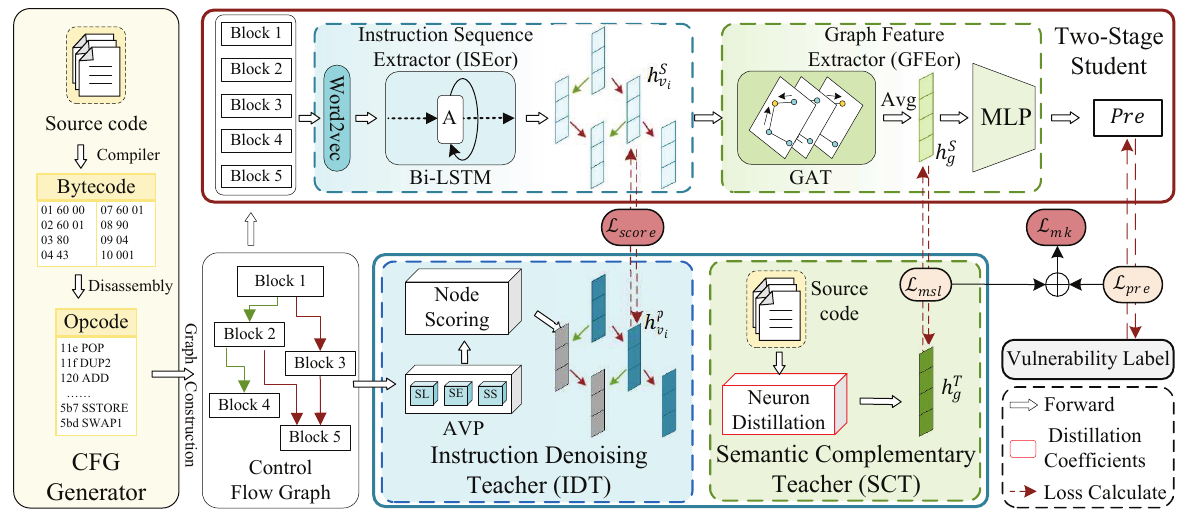}
	\caption{
		A high-level overview of MTVHunter. (1) CFG generator, which implements a transition between the source code and the Control Flow Graph. (2) Instruction denoising teacher, which eliminates the interference of noise in the CFG. (3) Semantic complementary teacher, which provides the missing semantic to bytecode by neuron distillation. (4) Two-stage student, which consists of a instruction sequence extractor and a graph feature extractor.
	}
	
	\label{fig:single_column_example}
\end{figure*}
\section{Preliminaries}
\textbf{Problem Definition.} Delivering a smart contract bytecode $ S_{byte}$, we are interested in developing a fully automated approach that can detect vulnerabilities only with bytecode. Specially, we train MTVHunter to predicts label \textit{y} $\in$ \{0,1\} of $ S_{byte}$ after being trained with the knowledge from two teachers. The ground truth $\mathcal{Y} = 1$  denotes $ S_{byte}$ has a certain type of vulnerability and $\mathcal{Y} = 0$ indicates $S_{byte}$ is benign. This paper focuses on four types of vulnerabilities.

\textit{Reentrancy vulnerability} is the classic and well-known vulnerability, the work \cite{a10} suggests that Reentrancy vulnerabilities have been always selected to study. It is triggered by the exploitation of gaps in debit and accounting operations. Specifically, when a user \textit{A} transfers money to another user \textit{B} via a function $f_1$, \textit{B}'s fallback function $f_b$ will be automatically triggered due to the default settings of the contract. At this point, $f_b$ becomes the key to illegal secondary transfers. The reason is that after $f_1$ transfers the money, the trigger of $f_b$ is able to enter $f_1$ again before \textit{A} has accounted it, thus illegally transferring A's money for many times, which ultimately leads to the occurrence of reentry.

\textit{Timestamp dependence vulnerability} is a type of malicious behavior that exploits the block timestamp to achieve unfair trading, i.e., if the timestamps are differ, the execution results of the contract will also differ. The reason for the difference is that the miner who mines the block is able to change the timestamp of the block within a short time interval (roughly 900 seconds) \cite{a7}. Thus, malicious miners may exploit this vulnerability by tampering with block timestamps to gain illegal profits.

\textit{TX.Origin vulnerability} is a type of malicious phishing behavior that uses the global variable tx.origin in a contract to track the original account address throughout its operation. If contract utilities tx.origin for user verification, it may be vulnerable to attackers, who could bypass the contract owner's verification and steal funds.

\textit{Delegatecall vulnerability} is a type of malicious action caused by exploiting the \textit{.Delegatecall()} method in solidity. The \textit{.Delegatecall()} function is originally intended to modify variables in the original contract while executing a function in the target contract. Attacker exploit this behavior to overwrite addresses in the original contract, bypassing address validation and illegally accessing users' funds.

\section{Method}
\subsection{Overview of the Methodology}
The overall framework of the proposed \textbf{M}ulti-\textbf{T}eacher \textbf{V}ulnerability \textbf{Hunter} (\textbf{MTVHunter}) is depicted in Fig. 1, consisting of four components: 1) a CFG generator, which constructs a control flow graph based on the dependency of the opcode dissembled from bytecode;
2) an instruction denoising teacher, which eliminates the interference of vulnerability non-relevant instruction (i.e., noise) via our proposed abstract vulnerability pattern, and reflects in node feature by node scoring mechanism;
3) a semantic complementary teacher, which creams valid feature off source code, and then maps them to bytecode missing semantic features via neuron distillation; 
and 4) a two-stage student, which absorbs the knowledge from these two teachers, and predicts whether the targeting smart contract is vulnerable or not. In the following, we elaborate on these four main components and discuss their core ideas. 
\subsection{CFG Generator}
As shown in Figure. 1, the CFG generator unfolds in three phases, i.e., compilation, disassembly, and graph construction. Concretely, we first employ solc\footnote{https://github.com/ethereum/solc-js } compiler to generate hexadecimal bytecode from source code, and then disassemble it into opcodes. Later, a CFG is constructed with the opcodes by an off-the-shelf symbolic execution solver, namely Octopus\footnote{https://github.com/FuzzingLabs/octopus/}. 
It is worth noting that the CFG is comprised of block nodes $ V = \{v_{1},v_{2},...,v_{n}\}$ and control flow edges $ E = \{ e_{ij}, i\neq j \}$, where the $ v_i$ is the i-th node, the \textit{n} denotes the total number of nodes, and the $e_{ij}$ represent the edge from node $ v_i$ to node $ v_j$. In general, each block node contains a set of instructions, while each edge represents the control relationship between nodes.

\subsection{Instruction Denoising Teacher }

To eliminate the noise interference on vulnerability detection, we present an instruction denoising teacher (IDT) with an abstract vulnerability pattern, which assists in identifying vulnerabilities for the student model.


\textbf{Abstract Vulnerability Pattern.} 
Inspired by existing local expert patterns \cite{a8}, we propose an abstract vulnerability pattern (AVP) based on the fact that the vulnerability occurrence is derived from the loading, operation, and storage of critical data. Compared with existing expert patterns, AVP is capable of highlighting vulnerability semantic buried under the noise instructions instead of being labor-consuming. 
Specifically, aiming at the potential vulnerability path, AVP defines three state-changing patterns to abstract the vulnerability semantics for smart contract, as shown in Table 1. 

Firstly, \textbf{State-Loading (\textit{SL})} targets at the invocation instructions, representing a critical data loading of smart contract. 
For example, CALLVALUE loads transaction data from the input for subsequent operations, and ORIGIN loads address data of the original account.
Secondly, \textbf{State-Operation (\textit{SO})} is designed to capture instructions that operate on the critical data after loading, such as arithmetic operation or logical operation.  
Lastly, \textbf{State-Storage (\textit{SS})} focuses on instructions used to record the changed data, which generally follows the first two patterns, including SSTORE, MSTORE and so on. 
\begin{table}[]
	\centering
	
	\begin{tabular}{c|l}
		\hline
		\textbf{Pattern}     & \textbf{Description}          \\ \hline
		\multirow{2}{*}{\textit{SL}} & Load the critical data, e.i.,CALLVALUE, \\
		& \textit{TIMESTAMP, TX.Origin and so on.}           \\ \hline
		\multirow{2}{*}{\textit{SE}} & Operations on loaded data. \textit{e.i., ADD,  }            \\  
		& \textit{GT, AND and so on.}            \\ \hline
		\multirow{2}{*}{\textit{SS}} & Stores the result of the operation.     \\  
		& \textit{e.i., SSTORE, MSTORE and so on.}    \\ \hline
	\end{tabular}
	\caption{Defined bytecode patterns and descriptions}
	\vspace{-2mm}
\end{table}

To further explain AVP, we take a snippet CFG of a \textit{withdraw} function as an example in Figure 2(a). 
The CALLVALUE in Block 1, matching \textit{SL}, indicates the loading of transaction amount. Then, the SUB in Block 3, conforming \textit{SE}, is assumed to be the operation of the amount. Finally, the SSTORE in Block 5, meeting with \textit{SS}, corroborates the results of the operation. Therefore, Block 1, Block 3, Block 5 are considered as a potential vulnerability path at the basic block level, while Block 2, Block 4 fail to match AVP and are identified as noise.

\begin{figure}[htbp]
	\centering
	\label{f2}
	
	\includegraphics[width=\linewidth]{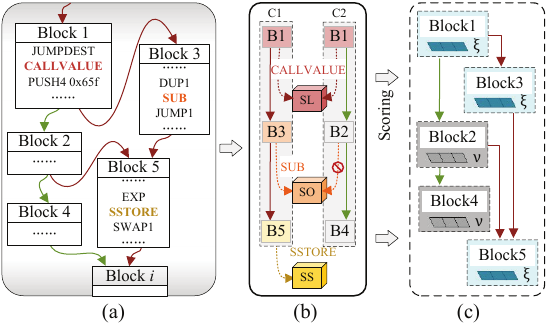}
	\caption{The overall process of IDT. (a) The snippet of CFG for vulnerable function \textit{Withdraw}. (b) The matching of AVP. (c) Node scoring Mechanism.  }
	\label{fig:single_column_example}
	
\end{figure}

\textbf{Node Scoring Mechanism.} 
With the help of AVP, we proposed a node scoring mechanism (NSM) that reflects denoising in node features. Since a control-flow path of CFG starts from a root node and ends at a leaf node, NSM constructs a node chain containing $l$ ( $l<6$) downstream nodes by depth first search (DFS) for each node. Then, it sequentially examines each node whether their instructions match the patterns of AVP, and assigns the matched nodes with score $\xi, \xi \in \text{(}0.5,1\text{]}$, while the remaining nodes are scored with $\nu,\nu \in\text{[}0,0.5\text{)}$. For instance, C1 and C2 are two node chains in Figure 4(b), which is extracted by NSM at $l=2$. The instructions within C1 satisfy three patterns of AVP, while C2 are not, the NSM scores $\xi$ for B1, B3, B5 and $\nu$ for B2, B4. 
Eventually, NSM embeds the scores to vector space as node features $h^{p}_{v_{i}}$ for the student to learn denoising knowledge.

\subsection{Semantic Complementary Teacher}
To tackle the challenge of restoring missing semantics for bytecode, we introduce a semantic complementary teacher (SCT), which leverages an effective and labor-saving neuron distillation mapping source code feature to the missing semantics of bytecode, and deliver to the student network for vulnerability detection.  

\textbf{Feature Extraction of Source Code.}
To capture effective information from source code, we present the smart contract as a graph and use temporal message passing (TMP) based on GNN for the graph embedding \cite{a9}, which is prevalent from the function control-flow graph representation. 
Particularly, it organizes the vulnerability-related functions as a core graph and captures core graph semantic features based on TMP.  Given a source code of smart contract, we exploit \textit{AutoExtractGraph} tool to generate a contract CFG and extract the graph semantic features $h_{g}^{sc}$ via TMP, which further feed to neuron distillation for obtaining the missing semantic of bytecode.

\textbf{Neuron Distillation.} After extracting the source code features, we propose a simple but effective method termed neuron distillation (ND), which maps the extracted source code features $h_{g}^{sc}$ to the missing semantic features $ h_{g}^{T}$ of bytecode. 
Specifically, ND adopts $N$ neurons as the distillation coefficient, as shown in Figure. 3, each neuron consists of a learnable matrix $ W = \{w_i, i\leq D\}$, a bias vector $b$ and an activation function $ \sigma$, where $D$ represents the dimension of the matrix. The features obtained by the neurons are followed with \textit{Average Pooling} to get missing semantic. The parameters of neurons are continuously adjusted with the training of the student network by multi-knowledge loss $L_{mk}$. Eventually, ND approximates the fitting of an optimal hyperplane for distillation. As following: 
\begin{equation}
	\setlength\abovedisplayskip{0pt}
	\setlength\belowdisplayskip{0pt}
	h_{g}^{T}=\frac{1}{N}\sum_{j=1}^{N}\sigma(\sum_{i=1}^{n}({w_{i}x_{i}+b}))
\end{equation}

\begin{figure}[htbp]
	\centering
	
	\label{f3}
	\includegraphics[width=\linewidth]{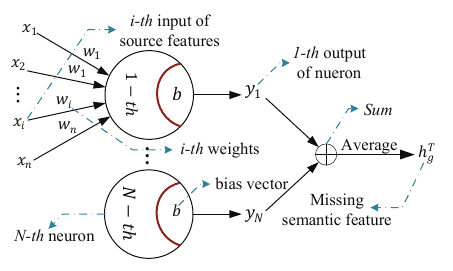}
	\caption{Illustration of neuron distillation mapping source code to missing semantic features. }
	\label{fig:single_column_example}
\end{figure}


By leveraging the neuron as the coefficient of distillation, ND turns the knowledge transfer into a regression task seeking the optimal coefficient. It distills valid knowledge from source code effectively to bytecode by an approximately optimal coefficient of distillation. Thus, ND dose not only effectively narrow the teacher-student gap, but also achieves a directional augmentation of features.

\subsection{Two-Stage Student}
To accurately detect the vulnerability for bytecode, we construct a two-stage student network, which consists of instruction sequence extractor and graph feature extractor, to absorb the knowledge from IDT and SCT.

\textbf{Instruction Sequence Extractor. }
For smart contract vulnerability detection, instruction sequence is essential as the main feature of contract execution path \cite{a13}, which usually gets obscured due to the noise.
As such, an instruction sequence extractor (ISEor) is designed to extract the bytecode sequence feature and simultaneously absorb the denoising knowledge. 
Specifically, ISEor first collects block nodes of the CFG in sequence and embeds each node to vector space $h_{v_i}^{w}$ via Word2vec. Then, it extracts the sequence feature $h_{v_i}^{S}$ hidden in the instruction sequence via Bi-LSTM $\mathcal{F}_{bi} $.
\begin{equation}
	\setlength\abovedisplayskip{0pt}
	\setlength\belowdisplayskip{0pt}
	h_{v_i}^{S}=\mathcal{F}_{bi}\left(h_{v_i}^{e}\right)
\end{equation}
To train ISEor, we compute the noise loss by using the mean-square error (MSE) loss between the node features $ h^{p}_{v_i}$ and the sequence feature $ h_{v_i}^{S}$.
\begin{equation}
	\setlength\abovedisplayskip{0pt}
	\setlength\belowdisplayskip{0pt}
	\mathop{min}\limits_{\hat{\theta}_{bi}} L(\hat{\theta}_{bi}) = \mathop{min}\limits_{\hat{\theta}_{bi}}\frac{1}{N}\sum_{i}(h_{v_i}^{p}-h_{v_i}^{S})^2
\end{equation}
where $ \hat{\theta}_{bi}$ denotes the parameters of Bi-LSTM, \textit{N} is the total number of nodes in the contract CFG. 

\textbf{Graph Feature Extractor. }
The bytecode of smart contracts is vested with incomplete control and data flow semantics due to miss a portion of semantics compared to the source code \cite{a7}. 
As such, we propose a graph feature extractor (GFEor), which extracts the globe graph feature from the contract CFG and assimilates the knowledge obtained from SCT to fill its missing semantics, to output the results of vulnerability prediction.
Concretely, we first reconstruct the CFG with sequence feature and control flow edges $ E = \{ e_{ij}, i\neq j \}$. Then, GFEor employs a graph attention network (GAT) $\mathcal{F}_{gat}$ to aggregate the information of neighborhoods and mine the globe graph features $h_{g}^{S} $ by average pooling. Thereafter, it predicts the label $y$ by Multilayer Perceptron (MLP). The formula is shown below:
\begin{equation}
	\setlength\abovedisplayskip{0pt}
	\setlength\belowdisplayskip{0pt}
	h_{g}^{S}=avgpool\left(\mathcal{F}_{gat}\left(h_{v_i}^{S},E\right)\right)
\end{equation}
\begin{equation}
	\setlength\abovedisplayskip{0pt}
	\setlength\belowdisplayskip{0pt}
	y=MLP\left(h_{g}^{S}\right)
\end{equation}

\textbf{Multi-Knowledge Loss.} To learn the missing semantic from ND, we propose a multi-knowledge loss to supervise the training of GFEor, and simultaneously train the neuron distillation coefficients. Specifically, multi-knowledge loss incorporates two signals of learning loss, one signal is label learning loss $L_{pre}$ calculated by the binary cross entropy (BCE) loss between the prediction $y$ and ground truth $ \mathcal{Y}$. The other is missing semantic learning loss $L_{msl}$ calculated by cross entropy (CE) between missing semantic features $ h_{g}^{T}$ and globe graph feature $h_{g}^{S} $. Given $N$ samples, the formula of the label learning loss and the complementary learning loss are shown in (6) and (7):
\begin{equation}
	\setlength\abovedisplayskip{0pt}
	\setlength\belowdisplayskip{0pt}
	L_{pre}=-\frac{1}{N} \sum \mathcal{Y} \log \left(y\right)+\left(1-\mathcal{Y}\right) \log \left(1-y\right)
\end{equation}
\begin{equation}
	\setlength\abovedisplayskip{0pt}
	\setlength\belowdisplayskip{0pt}
	L_{msl}=-\frac{1}{N}\sum h_{g}^{T} \cdot \log h_{g}^{S}
\end{equation}

Since $h_{g}^{T}$ and $h_{g}^{S}$ come from different modalities and are closely related to vulnerabilities, $L_{msl}$ inevitably receives a larger value \cite{b6}, resulting in greater impact to student and ND. Therefore, we introduce two hyper-parameters $\alpha$ and $\beta$ to balance the impact. The formula of the multi-knowledge loss is shown in (8).
\begin{equation}
	\setlength\abovedisplayskip{0pt}
	\setlength\belowdisplayskip{0pt}
	l_{mk} = \alpha \cdot L_{msl} +\beta \cdot L_{pre}
\end{equation}
Experiments show that $\alpha < \beta$ is most favorable for the learning of missing semantics.

\section{Experiments}

\begin{table*}[ht]
	\centering
	\renewcommand{\arraystretch}{1}
	\fontsize{8}{10}
	\selectfont 
	
	\scalebox{0.93}{
			\begin{tabular}{c|cccc|cccc|cccc|cccc}
				\hline
				\multirow{2}{*}{Methods} & \multicolumn{4}{c|}{Reentrancy} & \multicolumn{4}{c|}{Timestamp} & \multicolumn{4}{c|}{TX.Origin} & \multicolumn{4}{c}{Delegatecall} \\ \cline{2-17} 
				& Acc & Re & Pre & F1 & Acc & Re & Pre & F1 & Acc & Re & Pren & F1 & Acc & Re & Pre & F1 \\ \hline
				SmartCheck & 51.76&58.85& 52.40& 55.43
				&48.54&57.08& 39.03&46.36
				& 54.84&20.00&47.83& 28.21
				& 44.44&46.88& 53.57& 50.00\\
				Mythril & 63.14&52.9&75.61& 62.25
				& 64.94& 58.08&82.14& 68.05
				& n/a & n/a & n/a &n/a &61.11&84.21& 68.09&75.29\\
				ContractFuzzer & 64.51&71.5&52.27& 60.39
				& 58.77& 72.08&51.48& 60.06
				& n/a & n/a & n/a &n/a &70.37&36.36& 30.77&33.33
				\\
				Oyente &65.07  &63.02  &46.56  &53.55  &68.29  &57.97  &61.04  &59.47  & n/a &n/a  &n/a  & n/a & n/a & n/a &n/a  & n/a \\
				Securify & 68.63& 54.79& 45.98&50.00
				& n/a & n/a & n/a &n/a  &  n/a&n/a  &n/a  &  n/a& n/a &n/a  & n/a &n/a  \\
				Slither & 70.39& 66.92& 73.86& 70.22
				& 70.45& 41.40& 61.38& 49.44
				&n/a  & n/a & n/a &n/a  & 59.26&52.17& 52.17&52.17
				\\ 
				\hline
				Vanilla-RNN & 53.33 &32.52  & 40.36 & 36.02 &57.63  &14.77  & 14.10 &14.43  & 51.61 &47.37  &15.25  & 23.08 & 44.44 &18.75  & 60.00 & 28.57 \\
				Bi-LSTM &61.18  &71.26  &45.59  &55.61&68.99&85.00&51.35&64.03&69.35  &87.50  & 32.55 &59.57  &66.66  &82.35  & 70.00 &75.67 \\
				GAT & 80.00 &56.32&79.03&65.77&87.90&93.00&75.60&83.40 &88.71  & 78.12 & 78.12 & 78.12 & 79.62 &79.41  &87.09  &83.07  \\
				TMP &80.98&88.79&89.39&89.09
				&89.61&35.00&70.00&46.67
				&85.48&80.00& 66.67&72.73
				&88.89&85.00& 85.00&85.00\\
				SMS &81.90  & 90.28 & 67.81 & 77.69 &88.31  &89.5  &77.82  &83.25  &89.51  & 75.00 & 82.75 & 78.68 & 81.48 & 82.35 &87.50  &84.84  \\
				AME &84.12  & 79.61 &80.79 & 80.20 &83.12  &80.78  &84.64 &82.67 & 83.87 & 76.92 & 58.82 & 66.67 & 81.48 &80.00  &63.16 &70.59  \\
				DMT &86.86  & 92.00 & 75.23 & 82.77 &90.25  & 93.00 & 79.48 &85.71  & 91.12 & 87.5 & 80.00 & 83.58 &85.18  &82.35  & 93.33 &87.5  \\
				\hline
				\textbf{MTVHunter} &\textbf{95.10}&\textbf{100}&\textbf{87.44}  &\textbf{93.30}&\textbf{95.13}  &\textbf{99.50}  &\textbf{90.09}  &\textbf{92.72}  &\textbf{98.39} & \textbf{93.75} & \textbf{100} &\textbf{96.77}  &\textbf{88.89}  &\textbf{88.24}  &\textbf{93.75}  &\textbf{90.91} \\
				\hline
			\end{tabular}
		}
		\caption{Performance comparison (\%) in terms of accuracy (ACC), recall (RE), precision (PRE) and F1-score (F1). Fourteen methods are included in the comparisons. ‘n/a’ means the corresponding tool does not support detecting the vulnerability type.}
		
	\end{table*}
	
	\subsection{Experimental Setup}
	\textbf{Datasets.} We collected 229,178 public smart contracts from the official Ethereum website. 34,019 contracts are remaining after clearing (i.e., removing empty, repetitive, and simple ones). After that, we retained 10,356 contracts with valid compilations (except for compilations that take excessively long or that yield fewer functions). Further, we identify AVP for each contract to classify them into four vulnerability types. For example, contracts that possess \textit{SL} with CALLVALUE are susceptible to Reentrancy. Eventually, we manually labeled the ground truth in each category by auditing the source code of contracts, and split 1627 positive contracts and 5860 negative contracts.

	
	
	\subsection{Performance Comparison}
	
	\begin{table*}[ht]
		\centering
		
		\renewcommand{\arraystretch}{1}
		\scalebox{0.82}{
			\begin{tabular}{c|cccc|cccc|cccc|cccc}
				\hline
				\multirow{2}{*}{Variants} & \multicolumn{4}{c|}{Reentrancy} & \multicolumn{4}{c|}{Timestamp} & \multicolumn{4}{c|}{TX.Origin} & \multicolumn{4}{c}{Delegatecall} \\ \cline{2-17} 
				& Acc & Re & Pre & F1 & Acc & Re & Pre & F1 & Acc & Re & Pren & F1 & Acc & Re & Pre & F1 \\ \hline
				\textit{IDT-w/o} &72.55 &63.57  &78.1  &70.09&70.12&89.00&52.35&65.92&73.39  &87.10  & 79.41 &83.08  &72.22 &78.79  & 76.47 &77.61 \\ 
				\textit{IDT-PD }&80.00  & 56.89 & 78.57 & 66.00  &85.87  & 81.00 & 76.77 &78.83  & 85.48 & 80.00 & 66.67 & 72.73 &79.63  &84.85  & 82.35 &83.58 \\ 
				\textbf{\textit{IDT-CD}} &\textbf{90.98}  & \textbf{85.63} & \textbf{87.64} & \textbf{86.62} &\textbf{90.58}  & \textbf{72.00} & \textbf{98.63} &\textbf{83.23}& \textbf{87.10} & \textbf{72.22} & \textbf{81.25} & \textbf{76.47} &\textbf{83.33}  &\textbf{85.71}  & \textbf{82.76} &\textbf{84.21}  \\\hline
				\textit{SCT-w/o} &84.12  & 71.84 & 79.62 & 75.53 &83.28&78.84&79.17&79.00
				&79.03 &32.26 &66.67 & 43.48 &77.78  &88.24  &78.95 &83.33 \\
				\textit{1-Neuron }&90.98  & 85.63 & 87.64 & 86.62 &90.58  & 72.00 & 98.63 &83.23 & 87.10 & 40.91 & 75.00 & 52.94 &83.33  &85.71  & 82.76 &84.21 \\
				\textit{2-Neuron} &92.54  & 85.63 & 91.97 & 88.69 &89.44  & 96.00 & 77.10 &85.52  & 86.29 & 85.19 & 83.64 & 64.40 &87.04  &85.19  & 88.46 &86.79 \\ 
				\textit{\textbf{3-Neuron}} &\textbf{95.10}&\textbf{100}&\textbf{87.44}  &\textbf{93.30}&\textbf{95.13}  &\textbf{99.50}  &\textbf{90.09}  &\textbf{92.72}  &\textbf{98.39} & \textbf{93.75} & \textbf{100} &\textbf{96.77}  &\textbf{88.89}  &\textbf{88.24}  &\textbf{93.75}  &\textbf{90.91} \\\hline
			\end{tabular}
			
		}
		\caption{Performance comparison (\%) between MTVHunter and its variants on four vulnerabilities.}
	\end{table*}
	\textbf{Comparison with Traditional Detection Methods.}
	Following \cite{a12}'s strategy, we selected six SOTA traditional methods as benchmark tools, namely Oyente \cite{a4}, Securify \cite{a5}, Mythril \cite{a2} , ContractFuzzer \cite{a3}, Slither \cite{a6}, SmartCheck \cite{a1}. Considering the characteristics of these tools, we apply SmartCheck, Securify, and Slither for source-level vulnerability detection, and the other methods for bytecode-level vulnerability detection.
	In Table 2, we observe that the traditional tools have not yet achieved high accuracy on four vulnerabilities detection. For example, Slither yields a 70.39\% accuracy on Reentrancy vulnerability detection, which is attributed to the directly usage of simple and fixed expert patterns to detect vulnerabilities, resulting in high \textit{false-positive} and \textit{false-negative} rates.
	However, our proposed MTVHunter significantly achieves an accuracy of 95.10\%, gaining 24.71\% improvements over SOTA methods, owning to the appliance of AVP, which employs the expert patterns to achieving bytecode denoising.
	This suggests that the judicious application of expert patterns to denoising effectively contributes to the performance of smart contract vulnerability detection.
	
	\textbf{Comparison with Neural Network-Based Methods.} We selected three prevalent deep learning methods, i.e., Vanilla-RNN \cite{c1}, Bi-LSTM \cite{c2}, GAT \cite{c3}, and four open-source neural network-based SOTA detection tools, namely, TMP \cite{a9}, AME \cite{c4}, SMS, DMT \cite{a7} for a comprehensive evaluation. 
	To ensure fairness, bytecode sequences are fed to sequence-based Vanilla-RNN and Bi-LSTM, while the other GNN-based methods are presented with bytecode CFG.
	In Table 2, it is manifesting clear that MTVHunter performs significantly better than the sequence-based tools, and the accuracy gap has reached 26.14\% on Timestamp vulnerability detection. This is a result of sequence-based tools blindly employs bytecode sequences for vulnerability detection, which suffer from interference of noise. While MTVHunter eliminates the noise of bytecode sequence by AVP, clarifying the vulnerability information of smart contract. 
	Furthermore, DMT, the best GNN-based method, has a 86.86\% accuracy during TX.Origin vulnerability detection, while MTVHunter improves by 8.24\%. Technically, we speculate that the GNN-based methods may stem from two problems: 1) they are susceptible to noise when dealing with pure bytecode, increasing the difficulty of extracting vulnerability information. 2) these methods fail to restore the missing semantics of bytecode, which results in discontinuous semantics of the vulnerability.
	Our approach resolves the above issues and leads to impressive performance gains.
	
	\subsection{Ablation Study}
	\begin{table*}[ht]
		
		\centering
		
		\renewcommand{\arraystretch}{1}
		\scalebox{0.83}{
				\begin{tabular}{c|cccc|cccc|cccc|cccc}
					\hline
					\multirow{2}{*}{Methods} & \multicolumn{4}{c|}{Reentrancy} & \multicolumn{4}{c|}{Timestamp} & \multicolumn{4}{c|}{TX.Origin} & \multicolumn{4}{c}{Delegatecall} \\ \cline{2-17} 
					& Acc & Re & Pre & F1 & Acc & Re & Pre & F1 & Acc & Re & Pre & F1 & Acc & Re & Pre & F1 \\ \hline
					Vanilla-KD &50.58  & 58.04 & 36.07 & 44.49 &56.17& 17.60& 48.45&25.82
					& 52.42& 27.87& 53.13& 36.56
					&46.30&56.76& 61.76&59.15
					\\ 
					CTKD &60.58  &85.63  &45.84  &59.71 &62.01&20.60&49.48&29.09
					&54.84&27.78&46.88&34.88
					&66.67&94.44&50.00&65.38
					\\
					OFAKD &83.72  &92.52  &56.89  &70.46 
					&85.88 &87.03&83.88&85.43
					&80.65 &58.82&66.67&62.50
					&85.19 &73.68&82.35&77.78
					\\ 
					FT &88.03  &71.26  &91.85  &80.25 &85.23&53.33&49.48&51.34
					&83.06&62.79& 84.38&72.00
					&87.04&96.55& 82.35&88.89
					\\
					\hline
					\textbf{ND} &\textbf{95.10}&\textbf{100}&\textbf{87.44}  &\textbf{93.30}&\textbf{95.13}  &\textbf{99.50}  &\textbf{90.09}  &\textbf{92.72}  &\textbf{98.39} & \textbf{93.75} & \textbf{100} &\textbf{96.77}  &\textbf{88.89}  &\textbf{88.24}  &\textbf{93.75}  &\textbf{90.91} \\ \hline
				\end{tabular}
			}
			\caption{Performance comparison (\%) between neuron distillation and other existing distillation methods on four vulnerabilities.}
		\end{table*}

		\textbf{Instruction Denoising Teacher.}
		We first evaluate the effectiveness of IDT, with a focus on verifying the competence of AVP denoising, which is performed in the same configuration of SCT (ND with $N=1$). 
		In Table 3, MTVHunter yields the highest performance at complete denoising (IDT-CD with $\xi=1$,$\nu=0$), with an accuracy margin of 11.18\% in detecting Reentrancy vulnerability compared to without the IDT (IDT-w/o). 
		This shows that IDT indeed facilitates vulnerability detection of MTVHunter by eliminating the interference of noise. 
		However, we observe that partial denoising (IDT-PD with $\xi<1$, $\nu>0$) dose not always perform better than IDT-w/o. For instance, targeting the TX.Origin vulnerability, the accuracy of IDT-PD is higher than IDT-w/o, while the recall, precision and F1 value are lower. This occurrence indicates that IDT-PD is confused to discriminate between noise and vulnerability information, making the high \textit{false-positive }and \textit{false-negative} rates.
		In other words, IDT-PD may causes features that are originally vulnerability information to be judged as noise features after processing by the model. Therefore, the safeguarding of IDT effectiveness in vulnerability detection requires complete denoising of the noise to bytecode.
		
		\textbf{Semantic Complementary Teacher.}
		Subsequently, we evaluate the effectiveness of SCT, which is centered on validating the assistance provided by ND for student.
		In Table 3, the performance of MTVHunter improves with the increase of neurons and reaches the maximum F1 value at 3-Neuron, which is 93.3\%, 92.72\%, 96.77\%, 90.91\%, respectively. 
		This indicates that the optimal hyperplane fitted by multiple neurons yields a more comprehensive and effective bytecode semantic.
		Meanwhile, we observe that in detecting Timestamp vulnerability and TX.Origin vulnerability, the accuracy and precision are higher for 1-Neuron than 2-Neuron, while the recall and F1 value are lower. 
		We conjecture that the semantics obtained by the two neurons complement more vulnerability information, but are not stable enough to cause benign contracts to be misclassified as malicious.
		Later, we constructed 4-neurons to explore the stability of the semantics, and experiments revealed that the result of the four neurons are identical to that of the three neurons, exhibiting stable semantic.
		Hence, we are of the opinion that the collective work of several neurons ultimately yields an optimal missing semantic.

		\begin{figure*}
			
			\label{f4}
			\centering
			\includegraphics[width=\linewidth]{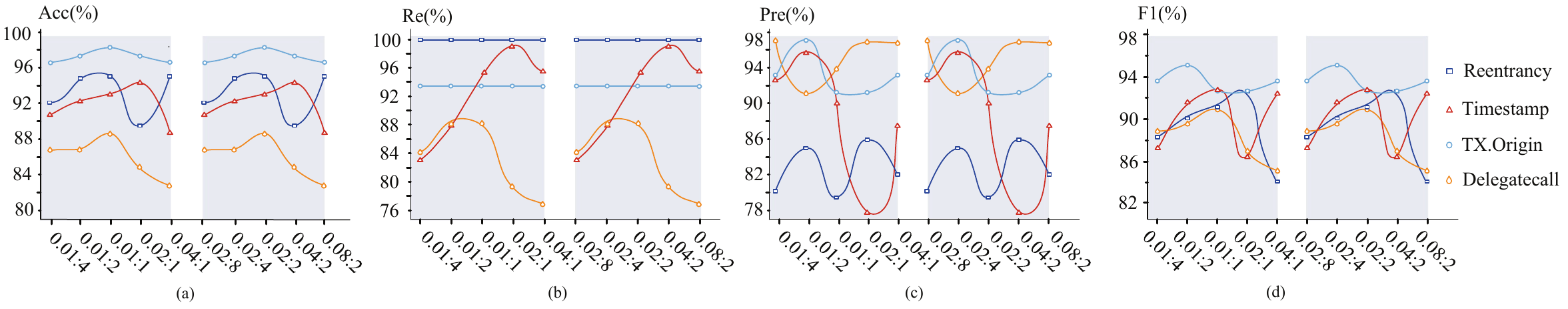}
			\caption{Performance comparison (\%) for different combinations of $\alpha_{sim}$ and $\beta_{pre}$ under the four vulnerabilities.}
			
		\end{figure*}
		
		On the other hand, multi-knowledge loss controls the training of ND while ND is the core component of SCT. Thus, we evaluate the contribution of multi-knowledge loss to SCT by adjusting $\alpha$ and $\beta$. 
		In Figure 4, we found that the performance of MTVHunter is identical when the ratio of $\alpha$ and $\beta$, denoted as $\delta$, is equal. 
		For example, MTVHunter simultaneously achieves 98.39\% accuracy in TX.Origin vulnerability detection with $0.01: 1$ and $0.02:2$. 
		We infer that the changes in $\alpha$ and $\beta$ do not affect the convergence of the model when the $\delta$ is the same.
		Meanwhile, Figure 4 shows that the F1 values of the four vulnerabilities behave differently as the ratio increases, with TX.Origin vulnerability undergoing the performance degradation earliest and the Reentrancy vulnerability the latest. We conclude that on account of TX.Origin vulnerability has less demand for missing semantics. When it overlearns the semantics, the performance degrades.
		In conclusion, multiple neurons and an appropriate ratio $\delta$ can contribute to a more complete semantic complement for student.

		\subsection{Comparison of Existing Distillation Methods}
		We selected four prevalent distillation methods, including Vanilla-KD \cite{b2}, CTKD \cite{b3}, OFAKD \cite{d1}, FT \cite{b5}, which are presented at top conferences and frequently used for comparison, to validate the effectiveness of neuron distillation (ND) at the same configuration,  As shown in Table 4, for Reentrancy vulnerability detection, the accuracy obtained by ND is 95.1\%, which outperforms Vanilla-KD and CTKD considerably. Compared to OFAKD and FT, ND also improves by 11.38\% and 7.07\%, respectively. This indicates that distillation methods with fixed or only one learnable distillation coefficient can be confusing for student networks to recover missing semantic information, while heterogeneous-aiming and encoding-based methods are relatively effective in restoring the missing semantic knowledge of bytecode, though they merely focus on recovering partial semantic information. 
		In contrast, the ND is more capable in recovering the complete semantics of bytecode by virtue of seeking optimal distillation coefficients.
		
		%
		
		\subsection{Interpretability Evaluation}
		
		We study the interpretability of the proposed AVP by visualizing the distribution of node features. 
		A vulnerable contract \textit{PoCGame} was harnessed as a case to study its node distribution under complete denoising.
		In Figure 5, there are 130 nodes, mapped to two-dimensional space by dimension by t-distributed Stochastic Neighbor Embedding (T-SNE), in which 15 red nodes meet the AVP and the remaining are noise nodes (green node). 
		
		As can be seen from the figure, on the one hand, the slope of the curve in Figure 5(b) is relatively larger than in Figure 5(a), which indicates that the aggregation of nodes in Figure 5(b) is more obvious. 
		Also, the distribution of green nodes show a larger variation while red nodes show a smaller variation, which directly reflects that the AVP alleviates the noise interference without mistakenly affecting critical nodes containing vulnerability semantics. 

		On the other hand, after the students learn the knowledge of denoising from IDT, we have the opportunity to compare the two node features of Figure. 5, certain values of the right node features are zero. In summary, the occurrence of 0 is directly related to $\nu=0$, implying that students are able to denoising autonomously.
		\begin{figure}[htbp]
			
			\centering
			\includegraphics[width=\linewidth]{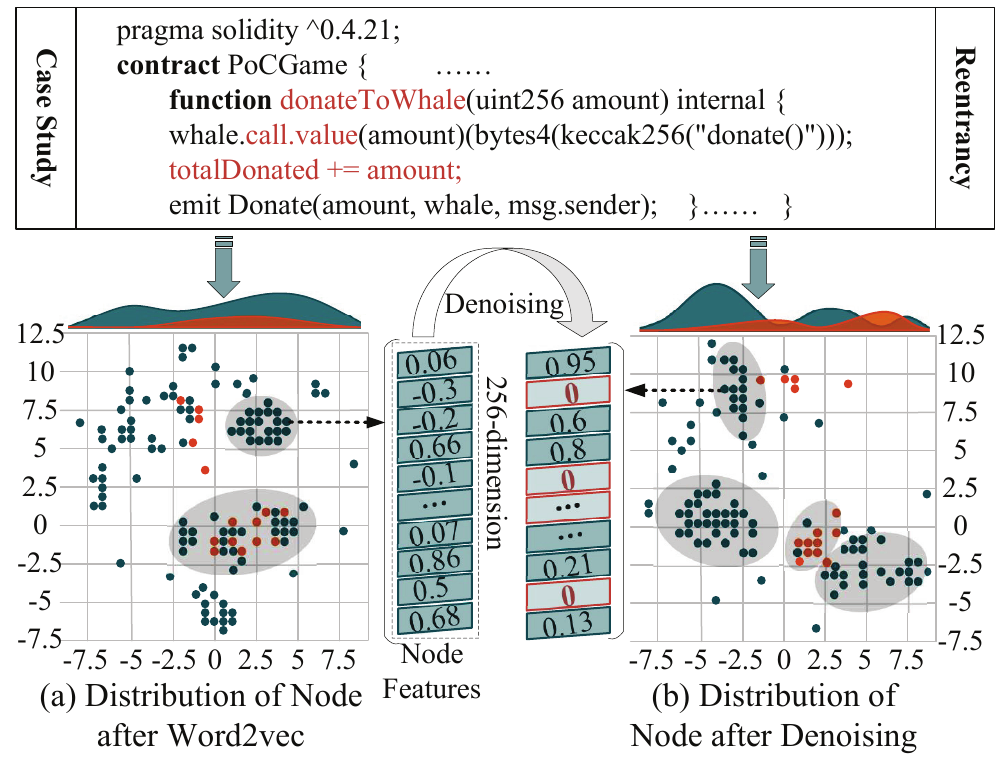}
			\caption{Case study on the interpretability of AVP. }
			\label{fig:single_column_example}
			
		\end{figure}
		\section{Relate Work}
		\subsection{Smart Contract Vulnerability Detection}
		Deep learning has been successful in various fields \cite{MM21_MotionPrediction} and
		most existing research employs them to detect vulnerabilities of smart contract. For example, 
		TMP \cite{a9} learns the temporal relationships of different program elements and aggregates them across the entire graph to detect vulnerabilities. AME\cite{c4} incorporates expert patterns into networks in an explainable fashion to automatically detect vulnerabilities. SMS and DMT\cite{a7} is based on mutual learning to detect the vulnerability of bytecode. 
		
		\subsection{Knowledge Distillation}
		Vanilla-KD \cite{b2} is the first knowledge distillation method, which is a response-based knowledge procedure.
		Factor Transfer \cite{b5} encodes the feature into a ‘factor’ using an auto-encoder to alleviate the leakage of information. CTKD \cite{b3} is a curriculum-based distillation method, introducing a learnable distillation coefficient for vanilla-KD. OFA-KD \cite{d1} is a model designed for distillation between heterogeneous architectures.

		\section{Conclusion}
		In this paper, we propose a fully automated smart contract bytecode vulnerability detection tool, MTVHunter. The approach achieves noise mitigation within instructions and restoring of missing semantic for bytecode. Compared to existing approaches, extensive experiments demonstrate that our proposed approach outperforms state-of-the-art methods and other neural network approaches. We believe that our work is an important step towards further highlighting the integration of deep learning methods with the vulnerability detection aspects of smart contracts.

\section{Acknowledgments}
This work is supported by  the National Key R\&D Program of China (No. 2023YFB3105904), the National Natural Science Foundation of China Youth Science Foundation Program (No. 62202121) and the Key R\&D Program of Zhejiang Province (No. 2023C01217). 

\bibliography{ref}

\end{document}